\documentclass{article}
\usepackage{spconf,amsmath,graphicx,color}
\usepackage{multirow}
\usepackage{url}
\usepackage{caption}
\usepackage{amsfonts}
\usepackage{cite}

\title{Supervised and Self-supervised Pretraining based COVID-19 detection using acoustic breathing/cough/speech signals}
\name{Xing-Yu Chen$^{1*}$, Qiu-Shi Zhu$^{1*}$, Jie Zhang$^{1,2}$, Li-Rong Dai$^{1}$
\thanks{*Equally contributed to this work.}
\thanks{This work was supported by the National Natural Science Foundation of China (No. 62101523),
Fundamental Research Funds for the Central Universities and the Leading Plan of CAS (XDC08010200). Thanks to the organizers of the DiCOVA-ICASSP 2022
Challenge for organizing this contest.}
\thanks{This paper presents a detailed introduction for the COVID-19 detection system from the team NELSLIP-USTC, which obtains the first place in tracks 3\&4 (i.e., speech \& fusion) of DiCOVA-ICASSP 2022 contest and the second place in track 1 (i.e., breathing).}}
\address{ $^1$NEL-SLIP, University of Science and Technology of China (USTC), Hefei, China\\
  $^2$State Key Laboratory of Acoustics, Institute of Acoustics, Chinese Academy of Sciences, Beijing, China}
\begin{document}
\ninept
\maketitle
\begin{abstract}
In this work, we propose a bi-directional long short-term memory (BiLSTM) network based COVID-19 detection method using breath/speech/cough signals. By using the acoustic signals to train the network, respectively, we can build individual models for three  tasks,  whose parameters are averaged to obtain an average model, which is then used as the initialization for the BiLSTM model training of each task. This initialization method can significantly improve the performance on the three tasks, which surpasses the official baseline results.
Besides, we also utilize a public pre-trained model wav2vec2.0 and pre-train it using the official DiCOVA datasets. This wav2vec2.0 model is utilized to extract high-level features of the sound as the model input to replace conventional mel-frequency cepstral coefficients (MFCC) features. Experimental results reveal that using high-level features together with MFCC features can improve the performance.
To further improve the performance, we also deploy some preprocessing techniques like silent  segment removal, amplitude normalization and time-frequency mask.
The proposed detection model is evaluated on the DiCOVA dataset and results show that our method achieves an area under curve (AUC) score of 88.44\% on blind test in the fusion track.
\end{abstract}
\begin{keywords}
COVID-19, binary classification, supervised pre-training, self-supervised pre-training, respiratory diagnosis.
\end{keywords}
\section{Introduction}
Since the outbreak of COVID-19, it quickly becomes pandemic all over the world, even at every conner.
This unknown disease has brought a serious influence to all countries ranging from global health, economy, education, trade, cultural exchange, etc.
Due to the fact that disease diagnosis is an important step for controlling  the transmission, it is worthy studying how to detect the COVID-19 rapidly and efficiently. Nucleic acid test is a common and traditional COVID-19 detection method, but patients have to be tested in specific place like hospitals, and the test is comparatively expensive and time-consuming.
With the development of artificial intelligence (AI) techniques in recent years, machine learning based methods have been frequently used in disease detection.
Compared to conventional medical diagnosis approaches, the AI-based methods have an obvious superiority, as the detection can be performed fully online and the time and economic expenses are much lower. The obtained diagnostic result can be used as  supplementary information for doctors to make a more accurate clinical discrimination.

It thus deserves to investigate {\it whether AI techniques can also be utilized for diagnosing COVID-19, where only medical images \cite{a9} or sounds \cite{a8} of patients will be recorded.} Recently, several AI-based techniques have been developed for the detection of COVID-19. For example, a ResNet50 based method was proposed in \cite{a1} based on the use of  computed tomography (CT) images, which achieves an accuracy of 96.23\%. Compared to the CT-based method, where the CT images are collected offline, it is more appealing if we can use sound signals (e.g., cough, speech, breath) to perform detection, as the latter can be recorded remotely and avoid the people gathering.

In principal, the detection of COVID-19 is a binary classification task (i.e., positive or negative), for which convectional machine learning methods can be leveraged.
In order to make use of sound data for classification, feature extraction is required as a pre-processing step. Mel-frequency cepstral coefficients (MFCC) and mel-frequency spectrogram that can reflect the nonlinear perceptive characteristics of human hearings to frequency are commonly used as features for sound activity analysis \cite{a10,a11}. In addition, zero crossing rate (ZCR), kurtosis, log energy, spectral centroid, roll-off frequency can also be used \cite{a5,a7}. In \cite{a2}, it was shown that the positive testee of COVID-19 have different acoustical parameters compared with the negative. In \cite{a3}, Coppock et al extracted the spectrogram feature and used a ResNet based CNN to detect cough sounds, which achieves an area under curve (AUC) of 0.846. In \cite{a4}, the support vector machine (SVM) was employed for the detection of COVID-19 in combination of voice signals and symptoms. In addition, other classifiers like long short-term memory (LSTM) \cite{a5}, k-Nearest Neighbor (kNN) \cite{a6}, Random Forest \cite{randomf} and Light gradient boosting machine (LightGBM) \cite{lgbm} can also be utilized in line with classical machine learning approaches.

As currently the amount of DiCOVA data is still limited, we  mainly aim at exploring whether additional pre-training can improve the detection performance in this work. For this, we explore a supervised pre-training method, an unsupervised pre-training method and ensemble the two methods using the official dataset. Specifically, for supervised pre-training, we utilize breath, cough and speech sounds to train  models, respectively, then we can obtain models for these three different tasks. The model parameters are averaged to obtain an average model. We treat the average model as an initialization model, which is taken to initialize the diagnosing model on different tasks. Experimental results show that this initialization method can significantly improve the performance on three tasks, which surpasses the official baseline results. For the unsupervised pre-training method, we pre-train the public pre-trained wav2vec2.0 model \cite{baevski2020wav2vec} on the DiCOVA dataset \cite{dicova2} and use the pre-trained model as a high-level feature extractor. We utilize the pre-trained model to extract high-level features on three tasks, which are input into the diagnosing model to replace the classic MFCC feature. Experimental results reveal that the developed unsupervised pre-training method also outperforms the official baseline, but it is worse than the supervised pre-training method. It is worth mentioning that ensembling the two methods can obtain a better result in the same task, which shows that ensemble different levels of information can improve the detection performance. More importantly, by fusing the output probabilities, we obtain the best performance in the fusion task.

\section{Methodology}
\label{sec:format}
\begin{figure}[!t]
  \centering
    \vspace{-0.65cm}
  \includegraphics[width=0.3\textwidth]{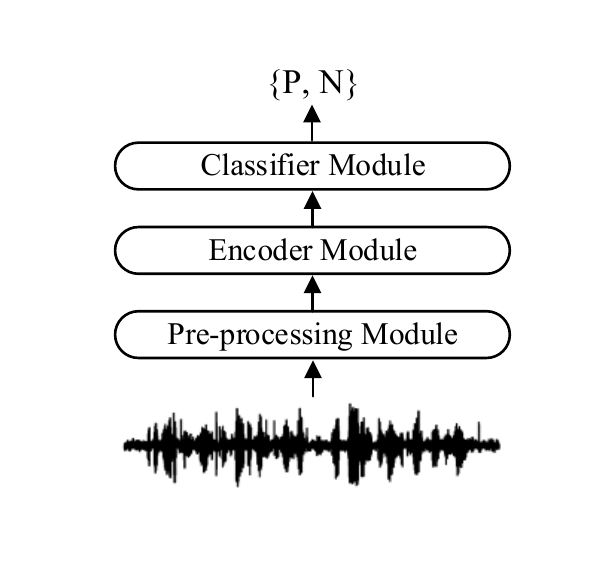}
  \vspace{-0.65cm}
  \caption{An illustration of the proposed DiCOVA diagnosing model.}
  \vspace{-0.3cm}
  \label{fig:dicova}
\end{figure}

\subsection{Model structure}
\label{sec:method}
The  diagram of the proposed classification model that we utilize in each subtask is shown in Fig.~\ref{fig:dicova}, which  mainly consists of a pre-processing module, an encoder module and a classifier module. In the encoder module, two bi-directional LSTM (BiLSTM) \cite{6795963} layers was utilized as the encoder. Each BiLSTM layer contains 128 hidden units and the dropout rate is set to be 0.1. Two fully connected feed-forward layers are utilized in the classifier module, which consists of two linear transformations with a ReLU activation in between. As the detection of COVID-19 is a 0-1 classification problem,  we use the binary classification loss function for training.

\subsection{Supervised pre-training method}
For the supervised pre-training, no extra data are utilized except the officially provided DiCOVA data. The total DiCOVA dataset consist of breathing, cough and speech labeled data. We obtain a breathing/cough/speech model using labeled breathing/cough/speech data through supervised pre-training, respectively. Specifically,  the breathing model $\text{Model}_{\rm breath}$ can be obtained using the labeled breathing data $X_{\rm breath}$ through supervised pre-training, and $\text{Model}_{\rm cough}$ and $\text{Model}_{\rm speech}$ can be built similarly as
\begin{align}
  \text{Model}_{\rm breath} &= \text{BiLSTM}(X_{\rm breath}),  \label{eq1}\\
  \text{Model}_{\rm cough} &= \text{BiLSTM}(X_{\rm cough}),  \label{eq2}\\
  \text{Model}_{\rm speech} &= \text{BiLSTM}(X_{\rm speech}).
  \label{eq3}
\end{align}

Then, we average the parameters of the three models as the average model as
\begin{equation}
  \text{Model}_{\rm ave} = {\rm ave}(\text{Model}_{\rm breath}, \text{Model}_{\rm cough}, \text{Model}_{\rm speech}).
  \label{eq4}
\end{equation}

\begin{figure}[!h]
  \centering
    \vspace{-0.65cm}
  \includegraphics[width=0.5\textwidth]{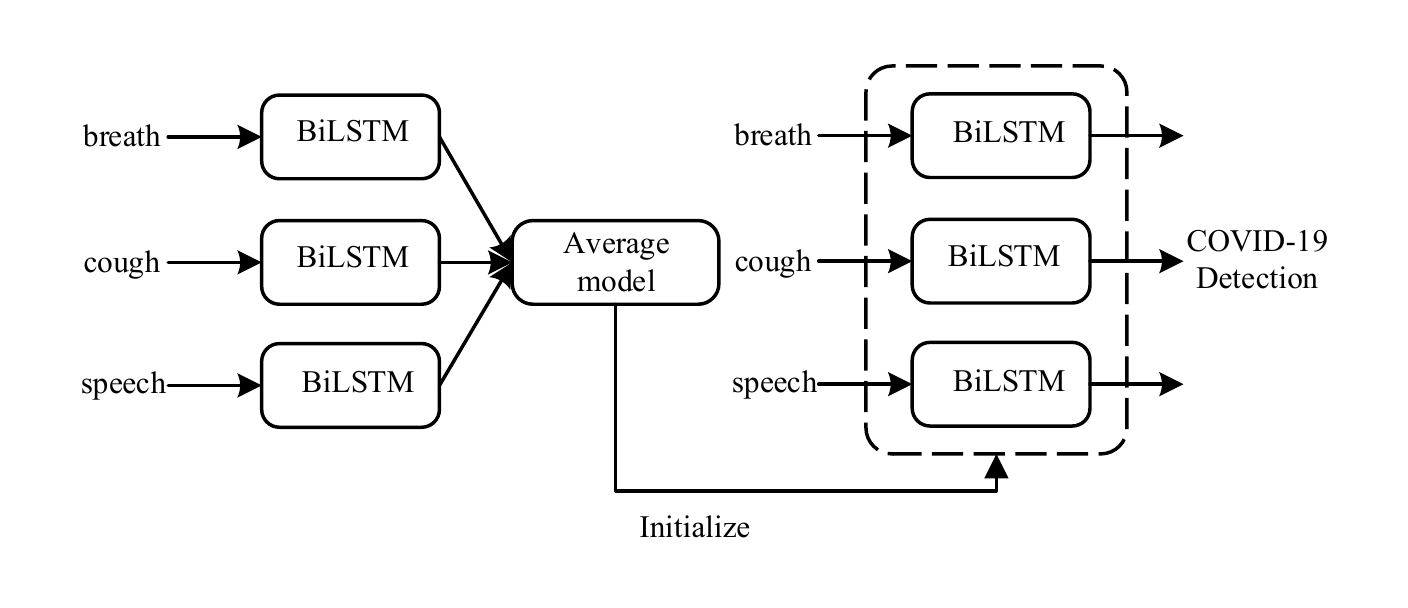}
  \vspace{-0.7cm}
  \caption{The supervised pre-training process.}
   \vspace{-0.2cm} 
  \label{fig:SupvsdTrn}
\end{figure}

For each task, the average model is considered as an initialization model, and we re-train the breathing/cough/speech model in the corresponding task. The pre-training process is shown in Fig.~\ref{fig:SupvsdTrn}. As we initialize the model using average model, the encoder and the classifier are easier to find the optimal solution.

\subsection{Self-supervised pre-training method}
As the amount of COVID-19 audio data in each sub-task is still limited, the traditional MFCC feature might be not sufficiently representative for classification tasks. Hence, we propose to utilize a high-level feature to replace the traditional MFCC feature. {It is worth mentioning that there are many effective methods that were proposed to solve these low-resource tasks~\cite{oord2018representation, chorowski2019unsupervised, Chung2019, steffen2019wav2vec, baevski2019vq}.} In the speech domain, The wav2vec2.0 \cite{baevski2020wav2vec} model is a representative self-supervised pre-training framework for learning speech representation. The public self-supervised pre-trained wav2vec2.0 \cite{baevski2020wav2vec} model is utilized in the stage of self-supervised pre-training. We pre-train the public pre-trained wav2vec2.0 model on the DiCOVA data and use the pre-trained model as a high-level feature extractor. For example, we use DiCOVA breathing data to pre-train the wav2vec2.0 model. After finishing pre-training, we utilize the pre-trained model to extract high-level breathing feature, which is then input into the diagnosing model to replace the MFCC feature.

\begin{figure}[!t]
  \centering
    \vspace{-0.4cm}
  \includegraphics[width=0.5\textwidth]{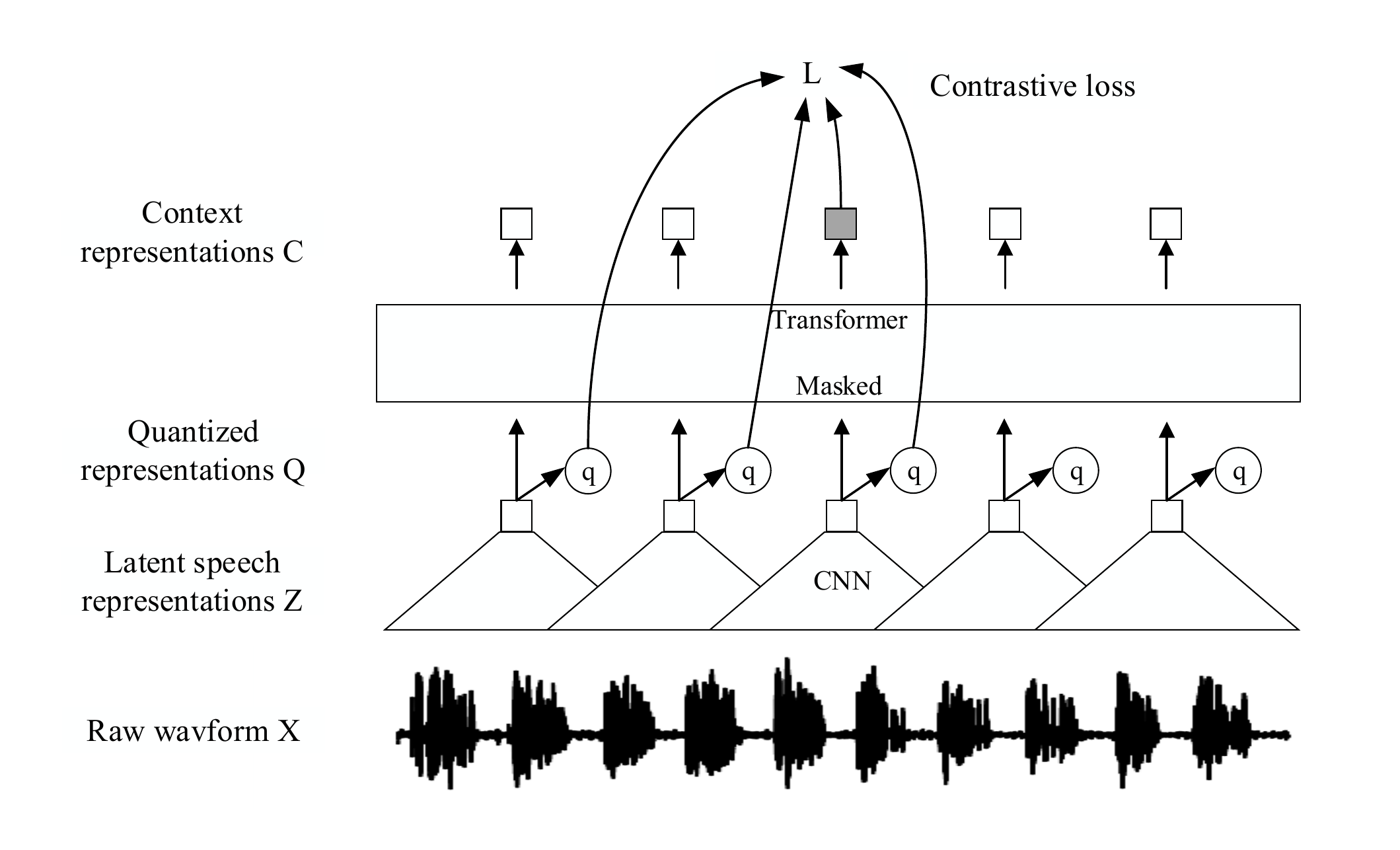}
  \vspace{-0.64cm}
  \caption{An illustration of the involved wav2vec 2.0 model.}
    \vspace{-0.3cm}
  \label{fig:wav2vec2}
\end{figure}

{In order to guide the reader, we will briefly review the wav2vec 2.0 model in this section. The structure of wav2vec 2.0 model is shown in Fig.~\ref{fig:wav2vec2}, including a CNN-based feature encoder $f:X \mapsto Z$ and a transformer encoder $g: Z \mapsto C $.  In detail, the input raw waveform $X$ is downsampled to the latent speech representation $Z$ by the feature encoder.} The transformer encoder then models the contextualized representation $C$ and {extracts a high-level feature} from the input $Z$. {A quantization module $Z \mapsto Q $ discretizes the output of the feature encoder to $q_t$}  as targets in the contrastive objective.

The quantization module first maps the latent speech representation $Z$ to logits $\mathbf{l} \in \mathbb{R}^{G \times V} $, given $G$ codebooks with $V$ entries.The Gumbel softmax operation \cite{jang2016categorical} is then used to select discrete codebook entries in a fully differentiable way. For a given frame $Z_t$, we can {therefore} select one entry from each codebook and concatenate the resulting vectors $e_1,...,e_G$ and apply a linear transformation to obtain $ \mathbf{q}_t $. The weighted loss function {is thus} given by
\begin{equation}
  L = L_{m}+\alpha L_{d} + \beta L_{f},
  \label{eq5}
\end{equation}
where
\begin{align}
  L_{m} &= -\log \frac{\exp({\rm sim}(\mathbf{c}_{t},\mathbf{q}_{t})/\kappa)}{\sum_{\tilde{\mathbf{q}} {\sim} \mathbf{Q}_t}\exp({\rm sim}(\mathbf{c}_t,\tilde{\mathbf{q}})/\kappa)},
  \label{eq6} \\
  L_{d} &= \frac{1}{GV}\sum_{g=1}^{G}\sum_{v=1}^{V}\overline{p}_{g,v}\log \overline{p}_{g,v},
  \label{eq7} \\
  \overline{p}_{g,v}&=\frac{\exp(\overline{l}_{g,v}+n_{v})/\tau}{\sum_{k=1}^{V}\exp(\overline{l}_{g,k}+n_k)/\tau}
  \label{eq8}.
\end{align}
It is clear that the total loss function is the weighted summation of three terms  $L_{m}$, $L_{d}$ and $L_{f}$ parameterized by $\alpha$ and $\beta$. In (\ref{eq6}), $L_{m}$ is the contrastive loss, which enables the model distinguishable between the true quantized latent speech representation $\mathbf{q}_{t}$ and a set of $K+1$ quantized candidate representations $\tilde{\mathbf{q}} \in \mathbf{Q}_{t}$. The quantized candidate representation $\tilde{\mathbf{q}}$ contains $\mathbf{q}_{t}$ and $K$ distractors, and the latter are uniformly sampled from other masked time steps of the same utterance. In (\ref{eq5}), the diversity loss $L_{d}$ aims to increase the use of quantized codebook representation, and $L_{f}$ is an $\ell_2$ penalty over the outputs of the feature encoder.  In (\ref{eq6}), sim stands for the cosine similarity between two vectors and $\kappa$ is a temperature. In (\ref{eq7}), $\overline{p}_{g,v}$ represents the probability of choosing the $v$-th codebook entry for group $g$ across a batch of utterances, where $\tau$ is a temperature. In (\ref{eq8}), $\overline{l}_{g,v}$ stands for the average logits $\mathbf{l}$ across utterances in a batch. More details on the  wav2vec 2.0 model can be found in \cite{baevski2020wav2vec}.

\subsection{Model ensemble method}

For the supervised pre-training method, the diagnosing model using MFCC feature as input, and for the self-supervised pre-training method, the diagnosing model using high-level features extracted by the pre-trained wav2vec2.0 model. As local features and high-level features contain different information, in order to combine the superiority of different features, we use model ensemble to improve the detection performance. Specifically, for the same task we utilize the proposed supervised  and self-supervised methods to train two models and obtain $\text{score}_{\rm sup}$ and $\text{score}_{\rm self-sup}$ on the test set, respectively. Then, we can ensemble the scores obtained by the two models, which is given by
\begin{equation}
  \text{score}_{\rm emsemble} = \mu * \text{score}_{\rm sup}+ (1-\mu) * \text{score}_{\rm self-sup},
  \label{eq9}
\end{equation}
where $\mu$ is chosen by the user. The weighted summation $\text{score}_{\rm emsemble}$ is taken as the detection probability.

For the track-4 fusion task, we utilize the best breathing, cough and speech models to obtain an approporiate fusion model as
\begin{equation}
  \text{score}_{\rm fusion} = \theta * \text{score}_{\rm bre}+ \gamma * \text{score}_{\rm cou} + \phi * \text{score}_{\rm spe} ,
  \label{eq10}
\end{equation}
where the parameter $\theta,\gamma,\phi $ need to satisfy $\theta+\gamma+\phi=1$. In principal, the breathing, cough and speech signals have a different importance in terms of the COVID-19 information, $\theta,\gamma,\phi $ should be chosen in line with the signal qualities. 

\section{Performance evaluation}
\subsection{Datasets}
The proposed model is evaluated on the DiCOVA-ICASSP 2022 challenge dataset \cite{dicova2}, which is derived from the crowd-sourced Coswara dataset \cite{coswara} and  collected from volunteers with different health conditions. Volunteers were advised to record their sound in a quiet environment using a web-application. The audio streams are sampled at a sampling frequency of 44.1 kHz  and in an FLAC format. The durations of the sound recordings range from about 1 second up to 29 seconds. The dataset consists of a total of 965 samples including 172 negative samples and 793 positive samples, where each sample includes cough, speech and breath sounds. Male and female patient samples are 723 and 242, respectively. The blind test set provided by the organizer includes 471 samples without labels.

\subsection{Model configuration}
In the pre-processing module, the amplitude of the raw waveform data is normalized between -1 to 1 through a normalization operator. As there are many silent segments in the speech/cough/breath sound signals, the speech activity detector (SAD) is applied to cut off these silent segments. The  sound data is downsampled to 16 kHz. Forty dimensional MFCC and delta-delta coefficients are extracted with a widow of 25 msec audio samples and a hop of 10 msec.  Due to the small size of the training data, we use SpecAugment \cite{specaug} time-frequency mask to augment the data, as it was shown data augmentation is effective to improve the performance, particularly in low-resource cases. The time mask length is 20 and the frequency mask length is 50. In the encoder module, two BiLSTM \cite{6795963} layers are utilized as the encoder. The BiLSTM layer dimension is 128 and the dropout rate is set to be 0.1. Two fully connected feed-forward layers are utilized in the classifier module, which consists of two linear transformations with a ReLU activation in between. The dimension of feed-forward layers is 256, which is finally mapped to 1-dimension for binary classification. As the detection of COVID-19 is a 0-1 classification problem,  we use the binary classification loss function for training. We follow the baseline system given by the organizer to train the model for 5-fold cross-validation in each task and then decode on the official blind test set to obtain the final test results, separately. In the baseline system, The CNN model is also utilized as another baseline system in experiments.

For the supervised pre-training, different seeds are set to train different models. The pre-training model is trained using 2 RTX3090ti-24G GPUs with 50 epochs, and the total training time is about 1 hour.
For the self-supervised pre-training, the wav2vec2.0 pre-training model is implemented using the fairseq toolkit \cite{ott2019fairseq}. The feature encoder contains seven blocks, where each block has 512 temporal convolution channels with strides (5, 2, 2, 2, 2, 2, 2) and kernel widths (10, 3, 3, 3, 3, 3, 2, 2). Thus, the
interval between two sequential samples in the feature encoder output $Z$ is around 20 ms and the receptive audio field is 25 ms. The models contain 12 transformer encoder blocks with a
dimension of 512, a feed forward module with a dimension of 2048 and 8 attention heads. The pre-training process is optimized with Adam \cite{kingma2014adam}. We set $G$ = 2  and $V$ = 320  for the quantization module and each entry with a size of 128. The temperature $\kappa$ is set to be 0.1 and  $\tau$ is annealed from 2 to 0.5 by a factor of 0.999995 over iterations. For the contrastive loss, $\alpha$ and $\beta$ are set to be 0.1 and 10, respectively. We use $K$ = 100 distractors and the total number of pre-training epochs are 200. After self-supervised pre-training, the pre-trained model parameters are frozen. Then, the pre-trained model is utilized as a high-level feature extractor, which generates high-level features from raw waveforms.

\begin{table}[!t]
\caption{ The AUC score of different methods on test/validation sets.}
\label{tab:table0}
\centering
\renewcommand\arraystretch{1.2}
\scalebox{1.0}{
\begin{tabular}{l|c|cc}
\hline
\multicolumn{1}{c|}{\multirow{2}{*}{\textbf{Method}}} & \multirow{2}{*}{\textbf{Model}} & \multicolumn{2}{c}{\textbf{ROC-AUC score (\%)}}           \\ \cline{3-4}
\multicolumn{1}{c|}{}                        &                        & \textbf{Test}                 & \textbf{Validation}           \\ \hline
\textbf{Track-1 breath}                               & \multicolumn{1}{l|}{}  & \multicolumn{1}{l}{} & \multicolumn{1}{l}{} \\
Official baseline                            & -                      & 84.50                & 77.63                \\
Baseline 1                                   & BiLSTM                 & 84.17                & 76.26                \\
Baseline 2                                   & CNN                    & 84.10                & 75.84                \\
Supervised pre-train                         & BiLSTM                 & 86.41                & 80.05                \\
Self-supervised pre-train                       & BiLSTM                 & 86.22                & 79.04                \\
Model ensemble                               & BiLSTM                 &\textbf{ 86.72}                & \textbf{80.05}                \\ \hline
\textbf{Track-2 cough}                                & \multicolumn{1}{l|}{}  & \multicolumn{1}{l}{} & \multicolumn{1}{l}{} \\
Official baseline                            & -                      & 74.89                & 75.88                \\
Baseline 1                                   & BiLSTM                 & 75.04                & 76.18                \\
Baseline 2                                   & CNN                    & 73.70                & 76.06                \\
Supervised pre-train                         & BiLSTM                 & 76.05                & 78.92                \\
Self-supervised pre-train                      & BiLSTM                 & 75.55                & 78.56                \\
Model ensemble                               & BiLSTM                 & \textbf{76.36}                & \textbf{78.92}                \\ \hline
\textbf{Track-3 speech}                                & \multicolumn{1}{l|}{}  & \multicolumn{1}{l}{} & \multicolumn{1}{l}{} \\
Official baseline                            & -                      & 84.26                & 82.24                \\
Baseline 1                                   & BiLSTM                 & 83.68                & 82.15                \\
Baseline 2                                   & CNN                    & 83.38                & 81.96                \\
Supervised pre-train                         & BiLSTM                 & 85.02                & 81.00                \\
Self-supervised pre-train                       & BiLSTM                 & 84.35                & 80.74                \\
Model ensemble                               & BiLSTM                 & \textbf{85.21}                & \textbf{81.30}                \\ \hline
\end{tabular}}
\end{table}

\begin{table}[!t]
\caption{ The AUC score of track-4 tasks on test and validation sets.}
\label{tab:table1}
\centering
\renewcommand\arraystretch{1.2}
\scalebox{1.0}{
\begin{tabular}{l|c|cc}
\hline
\multicolumn{1}{c|}{\multirow{2}{*}{\textbf{Track-4 fusion}}} & \multirow{2}{*}{\textbf{Fusion weight}} & \multicolumn{2}{c}{\textbf{ROC-AUC score (\%)}} \\ \cline{3-4}
\multicolumn{1}{c|}{}                                &                                & \textbf{Test        }  & \textbf{Validation}        \\ \hline
Official baseline                                    & -                              & 84.50         & 77.63             \\
Fusion1                                              & (1/3,1/3,1/3)                  & 87.01         & 82.56             \\
Fusion2                                              & (0.4,0.2,0.4)                  & \textbf{88.44}         & \textbf{82.93}             \\
Fusion3                                              & (0.5,0.1,0.4)                  & 88.44         & 82.71             \\ \hline
\end{tabular}}
\end{table}

\subsection{Results}
In experiments, we use 5-fold cross-validation to evaluate our supervised pre-training method, self-supervised pre-training method and model ensemble method. For comparison, we also test CNN, LSTM and official baseline model without pre-training. The official baseline comes from the website\footnote{https://competitions.codalab.org/competitions/34801\#results}. The obtained results are shown in Table~\ref{tab:table0} and Table~\ref{tab:table1}. Experimental results show that the methods with pre-training achieve a higher AUC than the baseline. That is, pre-training can provide more information for classifiers, particularly in case the dataset is small-sized.
Supervised pre-training models can achieve a great enhancement of performance compared with the baseline system. In breath/cough/speech task, supervised pre-training models increase the AUC from 84.50/74.89/84.26 to 86.41/76.05/85.02, respectively. Self-supervised pre-training methods obtain the AUC of 86.22/75.55/84.35 in the breath/cough/speech task. Despite achieving a smaller improvement in performance than supervised counterparts, self-supervised pre-training models still surpass the baseline in all tasks. Compared to the supervised and self-supervised methods, the proposed ensemble model can further improve the detection performance, which obtains the best performance  for all tasks, and the corresponding AUCs are 86.72/76.36/85.21 in cough/speech/breath tasks, respectively.
These results indicate that high-level feature pre-trained by wav2vec2.0 and low-level MFCC feature are complementary.
For the fusion task, the final probability is calculated as a weighted summation over three  best individual tasks. In case the cough/speech/breath weights are set to be 0.4/0.2/0.4, we achieve the highest AUC  of 88.44 for the fusion task.
\vspace{-0.2cm}
\section{Conclusion and discussion}
\vspace{-0.2cm}
In this work, we presented an ensemble model based on BiLSTM for diagnosing COVID-19 using acoustic signals. Due to the small size of the training data, we used supervised pre-traing and self-supervised pre-training methods, and more importantly both achieved a better performance than the baseline. Using the ensemble model of supervised pre-traing and self-supervised pre-training,  the AUC score was further improved, showing that the high-level feature pre-trained by wav2vec2.0 and low-level MFCC feature are complementary. The proposed model was evaluated on the DiCOVA challenge dataset and achieved an AUC score of 88.44\% in the blind test set for the fusion task, which reaches the first place in tracks 3\&4 of the DiCOVA-ICASSP 2022 contest.

\begin{figure}[!t]
  \centering
    \vspace{-0.2cm}
  \includegraphics[width=0.45\textwidth]{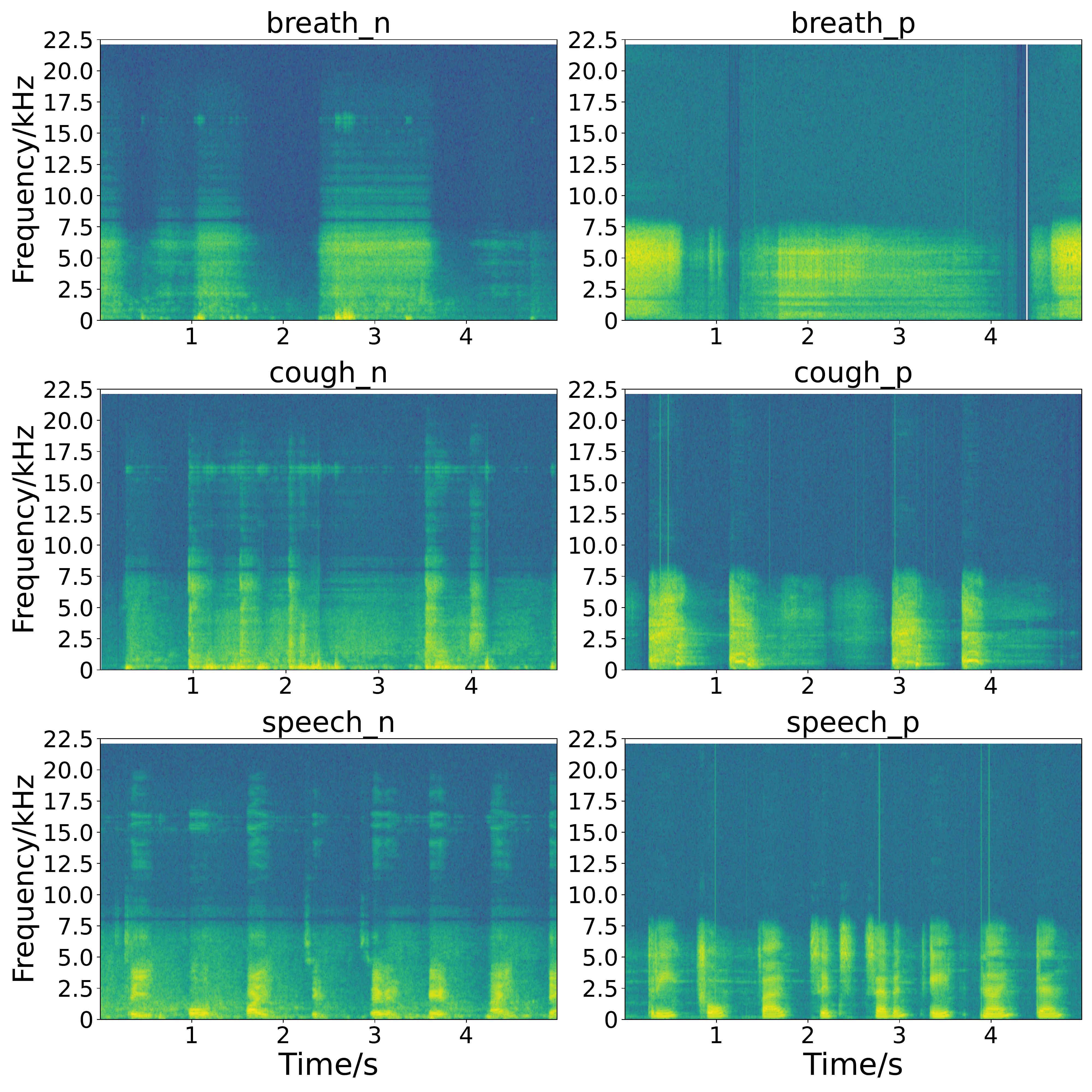}
  \vspace{-0.2cm}
  \caption{The spectrograms of the  positive (right) and negative (left) breathing/cough/speech audio samples.}
    \vspace{-0.4cm}
  \label{fig:spectrograms}
\end{figure}

As it was shown by experiments that in case of using the same classifier the high-level representation obtained by the wav2vec 2.0 model achieves a better detection performance than the MFCC feature, it tells that feature extraction is a vital step for the COVID-19 detection. In order to more clearly see the difference in the features of positive and negative audio samples, we show the spectrograms of the positive and negative breathing/cough/speech audio samples in Fig.~\ref{fig:spectrograms}. It is clear that for the positive samples, the energy of breathing/cough/speech signals is almost concentrated on low-frequency bands, while for negative samples the energy is distributed over full-frequency bands, particularly for the negative cough and speech signals. It reveals that high-level features are more beneficial for diagnosing COVID-19 using acoustic signals. In the future, we will therefore investigate an effective combination of the wav2vec 2.0 based acoustic representation, spectrograms, MFCC as a more representative feature for this detection problem.

\bibliographystyle{IEEEbib}
\bibliography{strings,refs}

\end{document}